\documentclass[twocolumn,showpacs,preprintnumbers,amsmath,amssymb]{revtex4}

\usepackage{graphicx}
\usepackage{dcolumn}
\usepackage{bm}

\newcommand{\bq}{\begin{equation}}
\newcommand{\eq}{\end{equation}}
\newcommand{\bqa}{\begin{eqnarray}}
\newcommand{\eqa}{\end{eqnarray}}
\newcommand{\nn}{\nonumber \\}

\def\be         {\begin{equation}}
\def\ee         {\end{equation}}
\def\bea        {\begin{eqnarray}}
\def\eea        {\end{eqnarray}}
\def\bnn        {\begin{eqnarray*}}
\def\enn        {\end{eqnarray*}}

\begin{document}

\title{Superconductivity in the presence of antiferromagnetism for high $T_c$ cuprates}
\author{Ki-Seok Kim, Sung-Sik Lee, Jae-Hyeon Eom and Sung-Ho Suck Salk$^{1}$}
\affiliation{Department of Physics, Pohang University of Science
and Technology Pohang 790-784, Korea \nn ${}^{1}$Korea Institute
of Advanced Studies, Seoul 130-012, Korea}
\date{\today}

\begin{abstract}
Using the U(1) holon pair slave boson theory [Phys. Rev. B {\bf
64}, 052501 (2001)], we derive a low energy field theory of
dealing with both $d-wave$ superconductivity and
antiferromagnetism for underdoped cuprates by constructing both
the Cooper pair field and the chargeon pair field. In terms of the
internal gauge field, the Cooper pair field carries no internal
charge while the chargeon pair field carries the internal charge.
They are decoupled in the low energy limit. This allows us to
separately treat the XY model of the Cooper pair field to describe
superconductivity and the Abelian Higgs model of the chargeon pair
field to describe antiferromagnetism in the presence of Dirac
fermions at and near the $d-wave$ nodal points. Thus we find that
the $d-wave$ superconductivity can coexist with antiferromagnetism
and that despite the coexistence, the antiferromagnetism can not
affect the superconducting transition, thus allowing the XY
universality class in the extreme type II limit.
\end{abstract}


\maketitle

\section{Introduction}

Recent experiments revealed anomalous weak antiferromagnetism in
underdoped superconducting
state\cite{Sonier_CSB,Keimer_CSB,Mook_CSB,Niedermayer_CSB,Julien_CSB}.
The coexistence of $d-wave$ superconductivity and
antiferromagnetism in underdoped cuprates has been examined by
various approaches (such as SO(5)\cite{Zhang_CSB},
stripes\cite{Kivelson_CSB,Zaanen_CSB,Sachdev_CSB},
QED3\cite{Franz_CSB}, $Z_2$ gauge theory\cite{YBKim_CSB}, SU(2)
slave-boson theory\cite{SU2_CSB}, mean field
theories\cite{Kyung_CSB,Wei-Min_CSB} and
etc.\cite{Moriya_CSB,Lichtenstein_CSB,Feng_CSB}). Thus it is of
great interest to answer whether the coexistence affects the
superconductivity.

In the context of U(1) slave boson theory Kim and
Lee\cite{DonKim_CSB} introduced the low energy Lagrangian in terms
of the massless Dirac spinon field $\psi_{l}$ and the massless
U(1) gauge field $a_{\mu}$ at half filling, \bqa &&{\cal L} =
\bar{\psi}_{l}\gamma_{\mu}(\partial_{\mu} - ia_{\mu})\psi_{l} .
\nonumber \eqa They showed that the $1/N$ perturbative treatment
leads to a dynamical mass generation (chiral symmetry
breaking\cite{CSB_DETAIL}) corresponding to an antiferromagnetic
ordering\cite{DonKim_CSB}. The above Lagrangian for a hole doped
system near half filling can be rewritten, \bqa &&{\cal L} =
\bar{\psi}_{l}\gamma_{\mu}(\partial_{\mu} - ia_{\mu})\psi_{l} +
\frac{K_b}{2}|\partial_{\mu}\phi_b - a_{\mu} - A_{\mu}|^2 .
\nonumber \eqa Here $K_{b}$ is the phase stiffness proportional to
hole concentration near half filling. $\phi_b$ is the phase field
of the holon, $a_{\mu}$, the internal U(1) gauge field and
$A_{\mu}$, the electromagnetic field. Based on this Lagrangian one
can show that in the dual field representation instantons
($(2+1)D$ U(1) magnetic monopoles) cause the single holon vortex
condensation\cite{NaLee_CSB}, thus implying phase incoherence of
the holon field. By integrating over the holon field in the
incoherent phase and redefining $a_{\mu} + A_{\mu}$ as
$a^{'}_{\mu}$, one obtains the low energy effective
Lagrangian\cite{Herbut_CSB}, \bqa &&{\cal L} =
\bar{\psi}_{l}\gamma_{\mu}(\partial_{\mu} - ia^{'}_{\mu} +
iA_{\mu})\psi_{l} + \frac{1}{2\bar{g}}|\partial\times{a}^{'}|^2 ,
\nonumber \eqa where $\bar{g}$ is the coupling strength, $\bar{g}
= 2|<\Phi_{b}>|^2$ with $|<\Phi_{b}>|$, the amplitude of the holon
vortex field. This Lagrangian is a QED Lagrangian in $(2+1)$
dimension. It is well known that chiral symmetry breaking occurs
for any nonzero coupling strength $\bar{g}$ as long as the flavor
number of fermions is less than a critical flavor number $N_c$
($N_c = \frac{32}{\pi^2}$)\cite{QED3_CSB}. Thus the
antiferromagnetic ordering at half filling is sustained near half
filling.

In this paper we examine the superconducting phase transition in
the low energy limit in the underdoped region. Earlier, the U(1)
and SU(2) holon-pair boson theory of Lee and
Salk\cite{LeeSalk_CSB} revealed the salient features of the
arch-shaped superconducting transition temperature in the phase
diagram of high $T_c$ cuprates and the peak-dip-hump structure of
optical conductivity in agreement with observations. It is of
interest to see whether the U(1) holon-pair boson theory of Lee
and Salk\cite{LeeSalk_CSB} reveals the coexistence of
antiferromagnetism (AF) with $d-wave$ superconductivity (SC) and
whether the antiferromagnetic order affects the $d-wave$
superconducting phase transition as a consequence of the
coexistence. In their U(1) holon-pair boson
theory\cite{LeeSalk_CSB}, the Cooper pair field is described as a
composite field of the spinon pair and holon pair fields. Applying
it to a low energy field theory, we first discuss how the
coexistence between the $d-wave$ SC and AF arises. We then
investigate whether the superconducting transition at $T = 0$
leads to the XY universality class independent of the
antiferromagnetic order. To facilitate this study composite
chargeon pair and Cooper pair fields are introduced. To avoid
confusion, we would like to point out that the chargeon pair field
here is different from the chargon field of the $Z_2$ gauge
theory\cite{SenthilFisher_CSB}. In our theory the internal U(1)
gauge field strongly couples to both the holon (charge $+e$ and
spin $0$) and the spinon (charge $0$ and spin $1/2$) and thus no
physical separation between the holon (charge) and the spinon
(spin) is allowed. As will be seen later our chargeon pair (which
is a composite particle made of a spinon pair and a holon pair)
has the internal U(1) charge $-4 \tilde e$ with the external
electromagnetic (EM) charge $-2e$. On the other hand, in the $Z_2$
gauge theory of Fisher and coworkers\cite{SenthilFisher_CSB}, the
chargon (spinon) is introduced in a different way. Their chargon
(spinon) is no longer the same as holon (spinon) in the U(1)
slave-boson theory. Instead the chargon (spinon) is constructed as
a composite of a holon (spinon) and a half of the spinon pair
phase field carrying no internal U(1) gauge charge and, thus, is
not coupled to the internal U(1) gauge field (likewise their
chargon pair carries no U(1) charge). Instead the chargon (spinon)
is minimally coupled only to the $Z_2$ gauge field. We also point
out that our composite vortex fields constructed from the spinon
pair and holon pair fields are physically different from those
introduced by Balents et. al\cite{Nayak_V_CSB,Nayak_CSB} in the
sense that their composite vortex fields are constructed from
spin-up and spin-down vortices.

In our composite vortex field approach the Lagrangian
(Eq.(\ref{eq8})) is divided into three sectors concerned with
Cooper pair vortices, chargeon pair vortices, and coupling between
the two vortices respectively. The key point of the present paper
is that the Cooper pair and chargeon pair sectors become decoupled
in the low energy limit despite the presence of the coupling
between the two sectors. As a result, in the low energy limit the
dual Lagrangian of the XY model describing superconductivity is
decoupled with the dual Lagrangian which describes
antiferromagnetism. From the decoupled Lagrangian, it is shown
that the AF can coexist with the SC without affecting the
superconducting transition. In our composite vortex field approach
the underdoped superconducting phase is seen to accompany the
condensation of the chargeon pair vortices in the absence of the
condensation of the Cooper pair vortices (Eq.(\ref{eq8}) and
Eq.(\ref{eq9})). Owing to free chargeon pair vortices (leading to
chargeon pair vortex condensation) the Berry gauge field
(Eq.(\ref{eq3})) coupled to the Dirac fermion remains massless
(Eq.(\ref{eq9})). Here the Berry gauge field is associated with
the Aharonov-Bohm phase acquired by the spinon when it moves
around a spinon pair vortex. Since the Berry gauge field
$a_\mu^\psi$ remains massless, the Lagrangian for the Dirac
fermion and the Berry gauge field reduces to the QED Lagrangian in
the low energy limit (${\bf \cal L}_{AF}$ in Eq.(\ref{eq10})).
This Lagrangian is in the same form as the Lagrangian of Kim and
Lee at half-filling\cite{DonKim_CSB}. The physical difference is
that the internal U(1) slave-boson gauge field $a_\mu$ at
half-filling is replaced by the Berry gauge field $a_\mu^\psi$
away from half-filling. The chiral symmetry breaking occurs even
away from half-filling as the case of the half-filling. This leads
to the antiferromagnetic order away from half-filling allowing the
coexistence with the superconductivity. On the other hand, we find
that the XY universality class of the $T = 0$ superconducting
phase transition in the extreme type II limit is not affected by
the presence of the antiferromagnetic order.

We would like to stress that our discussion on the problem of
coexistence between the antiferromagnetism and the
superconductivity differs from other
theories\cite{Franz_CSB,Zhang_CSB,SU2_CSB} in which competition or
interplay (correlation) between the antiferromagnetism and the
$d-wave$ superconductivity is emphasized. On the other hand, in
our theory we find that the superconductivity is not affected by
the antiferromagnetism, nor vice versa. Instead, the
antiferromagnetism is shown to have correlation with the Dirac
spinon field in association with the chargeon pair field, but not
with the Cooper pair field. Using our composite Higgs field
representation we describe the reason why the superconductivity is
not affected by the presence of antiferromagentic order in the
underdoped region.

\section{Construction of the low energy effective field theory and physical implications for $d-wave$ superconductivity and antiferromagnetism}

The $d-wave$ superconducting state emerges from antiferromagnetic
Mott insulating phase as a consequence of hole doping. At half
filling the ground state shows an antiferromagnetic long-range
order. As hole doping increases, the AF fluctuations may accompany
the $d-wave$ spin singlet pair excitations in the pseudogap (PG)
phase. When the spin singlet pairs are available, doped holes
prefer to be paired at adjacent sites since such configurations
are energetically favored to allow for mutual hopping without bond
breaking of the pairs. In the context of the slave boson theory an
empty site can be regarded as a holon, i.e., an object of charge
$+e$ and spin $0$. In the PG phase the holon pairs remain
phase-incoherent and thus they are the preformed pairs. When the
preformed holon pairs are bose-condensed, $d-wave$
superconductivity may arise as a result of forming the Cooper pair
as a composite of the spinon pair of $d-wave$ symmetry and the
holon pair of $s-wave$ symmetry. This picture is well addressed in
the U(1) slave boson Hamiltonian of Lee and
Salk\cite{LeeSalk_CSB}, \bqa &&H^{eff}= H^{eff}_{0} + H^{eff}_{s}
+ H^{eff}_{b} , \nn &&H^{eff}_{0} = \sum_{<i,j>}\Bigl[ J_{\delta}
\Bigl( \frac{1}{2}|\Delta_{ji}^{sp}|^{2} + \frac{1}{4}
|\chi_{ji}|^{2} + \frac{1}{4}\Bigr) \nn && +
\frac{J}{2}|\Delta^{sp}_{ij}|^2\Bigl(|\Delta_{ji}^{bp}|^{2} +
\delta^2 \Bigr) \Bigr] , \nn &&H^{eff}_{s}= -\frac{J_{\delta}}{4}
\sum_{<i,j>,\sigma} \Bigl[ \chi_{ji}^{*}
(f_{j\sigma}^{\dagger}f_{i\sigma}) + c.c. \Bigr] - \sum_{i,\sigma}
\mu^{f}_{i}(f_{i\sigma}^{\dagger} f_{i\sigma}) \nn &&
-\frac{J_{\delta}}{2} \sum_{<i,j>} \Bigl[ \Delta_{ji}^{sp*}
(f_{j\uparrow}f_{i\downarrow}-f_{j\downarrow}f_{i\uparrow}) + c.c.
\Bigr] , \nn &&H^{eff}_{b} = -t \sum_{<i,j>} \Bigl[
\chi_{ji}^{*}(b_{j}^{\dagger}b_{i}) + c.c.  \Bigr] -\sum_{i}
\mu_{i}^{b} ( b_{i}^{\dagger}b_{i} ) \nn && -\sum_{<i,j>}
\frac{J}{2}|\Delta^{sp}_{ij}|^2 \Bigl[ \Delta_{ji}^{bp*}
(b_{i}b_{j}) + c.c. \Bigr] . \label{eq1} \eqa Here $J_{\delta} =
J(1-\delta)^2$ is the doping ($\delta$) dependent renormalized
Heisenberg coupling strength. $\chi_{ji}= <
f_{j\sigma}^{\dagger}f_{i\sigma} + \frac{4t}{J(1-\delta)^2}
b_{j}^{\dagger}b_{i}>= \chi_{ji}^s + \frac{4t}{J(1-\delta)^2}
\chi_{ji}^b$ is the effective hopping order parameter in the
uniform phase with $\chi_{ji}^s  = <
f_{j\sigma}^{\dagger}f_{i\sigma}>$, that of the spinon, and
$\chi_{ji}^b = < b_{j}^{\dagger}b_{i}>$, that of the holon.
$\Delta_{ji}^{sp}=<
f_{j\uparrow}f_{i\downarrow}-f_{j\downarrow}f_{i\uparrow} >$ is
the $d-wave$ spinon singlet pairing order parameter and
$\Delta_{ji}^{bp} = <b_{j}b_{i}>$, the $s-wave$ holon pairing
order parameter. The main physics is imposed by the above
effective Hamiltonian (the last term, $-\sum_{<i,j>}
\frac{J}{2}|\Delta^{sp}_{ij}|^2 \Bigl[ \Delta_{ji}^{bp*}
(b_{i}b_{j}) + c.c. \Bigr]$ in Eq.(\ref{eq1})) which shows
coupling between the spinon pairing and the holon pairing orders.
Judging from the coupling term, we expect that the holon-pair bose
condensation can emerge only in the presence of spinon pairing.

In order to construct a low energy effective Lagrangian from
Eq.(\ref{eq1}), we first pay attention to the spinon Hamiltonian.
Considering the low energy excitations we take only the phase
fluctuations of the spinon pairing order parameter. It is of note
that the phase of the spinon pairing order parameter is coupled to
the spinon in the pairing interaction term in Eq.(\ref{eq1}), that
is, $ \pm \frac{J_{\delta}}{2} \sum_{<i,j>} |\Delta^{sp}_{ij}|
\Bigl[ e^{-i\phi^{sp}_{ij}}
(f_{j\uparrow}f_{i\downarrow}-f_{j\downarrow}f_{i\uparrow}) + c.c.
\Bigr]$. Here $\phi^{sp}_{ij}$ ($|\Delta^{sp}_{ij}|$) is the phase
(amplitude) of the spinon pairing order parameter. $+1$ ($-1$) for
the phase factor is for $ij$ link parallel to the $\hat x$ ($\hat
y$) direction representing the d-wave pairing symmetry of the
spinon pairing order parameter. In order to take care of the
coupling between the vortex of the spinon pair and the spinon
quasiparticle, we consider a singular gauge
transformation\cite{Herbut_CSB,Tesanovic_CSB,Ye_CSB} for the
spinon fields, $\psi_{i\uparrow} =
e^{-i\phi_{\uparrow{i}}}f_{i\uparrow}$  and $\psi_{i\downarrow} =
e^{-i\phi_{\downarrow{i}}}f_{i\downarrow}$ with $\phi_{ij}^{sp} =
\phi_{\uparrow{i}} + \phi_{\downarrow{j}} = \phi_{\downarrow{i}} +
\phi_{\uparrow{j}}$. As a consequence, the two phase fields appear
in the kinetic energy term for the renormalized spinon
$\psi_{i\sigma}$; one as a Berry gauge field $a_{ij}^{\psi} =
\frac{1}{2}(\phi_{\uparrow{j}} - \phi_{\uparrow{i}}) -
\frac{1}{2}(\phi_{\downarrow{j}} - \phi_{\downarrow{i}})$ and the
other as a Doppler gauge field $v_{ij} =
\frac{1}{2}(\phi_{\uparrow{j}} - \phi_{\uparrow{i}}) +
\frac{1}{2}(\phi_{\downarrow{j}} -
\phi_{\downarrow{i}})$\cite{Herbut_CSB,Tesanovic_CSB,Ye_CSB}. The
Berry gauge field causes the renormalized spinon to acquire the
Aharonov-Bohm phase when it moves around a spinon pair vortex. The
Doppler gauge field causes a shift in the energy of the
renormalized spinon in the presence of the spinon pair current.
With the renormalized spinon fields $\psi_{i\sigma}$, the spinon
Hamiltonian $H_s^{eff}$ in Eq.(\ref{eq1}) is rewritten as \bqa
&&H^{eff}_{s}= -\frac{J_\delta}{4} \chi \sum_{<i,j>}
\Bigl(\psi_{i\uparrow}^{\dagger}  e^{-i(a_{ij}- v_{ij} -
a_{ij}^{\psi})} \psi_{j\uparrow} \nn && +
\psi_{i\downarrow}^{\dagger}e^{-i(a_{ij}- v_{ij} + a_{ij}^{\psi})}
\psi_{j\downarrow} + h.c. \Bigr) \nn && + \frac{J_\delta}{2}
\Delta_{sp} \sum_{i} \Bigl( (\psi_{i\uparrow} \psi_{i+{\hat
x}\downarrow} - \psi_{i\downarrow} \psi_{i+{\hat x}\uparrow}) +
h.c. \Bigr) \nn && - \frac{J_\delta}{2} \Delta_{sp} \sum_{i}
\Bigl( (\psi_{i\uparrow} \psi_{i+{\hat y}\downarrow} -
\psi_{i\downarrow} \psi_{i+{\hat y}\uparrow}) + h.c. \Bigr) \nn &&
- \sum_{i,\sigma} \mu^{f}_{i} (\psi_{i\sigma}^{\dagger}
\psi_{i\sigma}). \label{eq1-1} \eqa The internal U(1) gauge field
$a_{ij}$ arises as a result of hopping of spinons (or holons)
around plaquettes. Here we use the uniform amplitudes for the
hopping order parameter ($\chi$) and the spinon pairing order
parameter ($\Delta_{sp}$) neglecting the amplitude fluctuations of
the order parameters in the low energy.

We now take the continuum limit for both the fermion field
$\psi_{i\sigma}$ and the phase of the spinon pairing order
parameter $\phi^{sp}_{ij}$ in the low energy. The spinon
excitations near the d-wave nodal point are described by the Dirac
Lagrangian in the continuum limit. The Dirac fermion is coupled to
the Berry gauge field $a_\mu^\psi$, the Doppler gauge field
$v_\mu$ and the internal U(1) gauge field $a_\mu$. The low energy
Lagrangian for the Dirac fermion coupled to the three gauge fields
are obtained to be, from Eq.(\ref{eq1-1})\cite{Ye_CSB}, \bqa {\bf
\cal L}_{\psi} &= & \bar\psi_l\gamma_\mu \partial_\mu \psi_l   -2
J_{c\mu} ( v_\mu - a_\mu ) - J_{s\mu} a^\psi_\mu \nn
 &= & \bar\psi_l\gamma_\mu \partial_\mu \psi_l   - J_{c\mu} ( \partial_\mu
\phi_{sp} - 2 a_\mu  ) - J_{s\mu} a^\psi_\mu. \eqa Here $\psi_l$
represents the 4 component spinor of the massless Dirac fermion
near the nodal points (l = 1 and 2), $\gamma_\mu$, the Dirac
matrices in (2+1) dimension and $\tau^i$, the Pauli
matrices\cite{Ye_CSB}. The Berry gauge field and the Doppler gauge
field become $a^{\psi}_{\mu} = \frac{1}{2} ( \partial_{\mu}
\phi_{\uparrow} - \partial_{\mu} \phi_{\downarrow} )$ and $v_{\mu}
= \frac{1}{2} ( \partial_{\mu} \phi_{\uparrow} + \partial_{\mu}
\phi_{\downarrow} ) = \frac{1}{2}\partial_{\mu} \phi_{sp}$ (note
that we replaced $\phi^{sp}_{ij}$ with $\phi_{sp}$ for the phase
of the spinon pairing order parameter) respectively in the
continuum limit. $J_{s\mu} = i( \bar\psi_l\gamma_0 \psi_l, v_f
\bar\psi_l\gamma_1 \psi_l  , v_\Delta \bar\psi_l\gamma_2 \psi_l )$
is the spin current of the Dirac fermion and $J_{c\mu} = (
\sum_{l} \psi_{l}^\dagger\tau^3\psi_{l}, v_f
\psi_{1}^\dagger\psi_{1}, v_f \psi_{2}^\dagger\psi_{2})$, the
internal U(1) charge ($q = \tilde e$) current of the Dirac fermion
($\tilde{e}$ denotes unit internal U(1) charge). $v_f \sim
J_\delta \chi$ and $v_\Delta \sim J_\delta \Delta_{sp}$ is the
Fermi velocity and the gap velocity respectively\cite{Ye_CSB}. In
the following we will set $v_f = v_\Delta = 1$ for simplicity.

We are interested in the construction of an effective Lagrangian
involved with low energy excitations. Following the U(1) slave
boson Hamiltonian introduced above, the low energy excitations
refer to the phase fluctuations (but not the amplitude
fluctuations) of the spinon pair and holon pair order parameters,
the massless spinon (Dirac particle) excitations (at the $d-wave$
nodal points), and the single holons. Gauge field fluctuations
$a_\mu(x)$ are introduced to allow the presence of internal flux
responsible for energy lowering, which arises as a result of
hopping of spinons (or holons) around plaquettes. Another low
energy excitation is concerned with the Berry gauge field
$a_{\mu}^\psi$\cite{Herbut_CSB,Tesanovic_CSB,Ye_CSB} minimally
coupled to the massless Dirac fermions. A singular gauge
transformation is considered to allow for the formation of spinon
pair vortices. Thus considering the above elementary excitations,
we write the $(2+1)D$ low energy Lagrangian, \bqa &&Z = \int
{D\psi_l}{D\psi_b}{D\phi_{sp}}{D\phi_{bp}}{Da_{\mu}^\psi}{Da_{\mu}}{D\lambda_{\mu}}
e^{-\int{d^3x} {\cal L} } , \nn &&{\cal L} = {\cal L}_{s} + {\cal
L}_{b}, \nn &&{\cal L}_{s} =
\frac{K_{sp}}{2}|\partial_{\mu}\phi_{sp} - 2a_\mu -
K_{sp}^{-1}J_{c\mu}|^2 \nn && +
i\bar{\rho}_{sp}\partial_{\tau}\phi_{sp} +
\bar\psi_l\gamma_\mu(\partial_\mu - ia_{\mu}^\psi)\psi_l + i
\lambda_{\mu} ( \partial \times a^{\psi} - \frac{1}{2} j_s^V
)_{\mu} , \nn &&{\cal L}_{b} =
\frac{K_{bp}}{2}|\partial_{\mu}\phi_{bp} - 2a_\mu - 2A_\mu -
K_{bp}^{-1}J_{b\mu}|^2 \nn && +
i\bar{\rho}_{bp}(\partial_{\tau}\phi_{bp} - 2A_0) +
|\partial_{\mu}\psi_b|^2 . \label{eq2} \eqa In the above equation,
$\phi_{sp}$ and $\phi_{bp}$ are the phase (not the phase factor)
of the spinon pair and holon pair order parameters respectively.
$K_{sp} \sim J_{\delta}|\Delta_{sp}|^2$ is the phase stiffness of
the spinon pair order parameter and $K_{bp} \sim
J|\Delta_{sp}|^2|\Delta_{bp}|^2$, the phase stiffness of the holon
pair order parameter. $\psi_b = e^{-i\frac{1}{2}\phi_{bp}}b$ is
the renormalized holon quasiparticle with $b$, the bare holon and
$J_{b\mu} = -i(\psi_{b}^{\dagger}\partial_{\mu}\psi_{b} -
\psi_{b}\partial_{\mu}\psi_{b}^{\dagger})$, the holon
quasiparticle three current. The coupling between the spinon
(holon) current $J_{c\mu}$ ($J_{b\mu}$) and the gauge invariant
spinon (holon) pair current $\partial_{\mu}\phi_{sp} - 2a_\mu$
($\partial_{\mu}\phi_{bp} - 2a_\mu- 2A_\mu$) in the first term of
${\cal L}_{s}$ (${\cal L}_b$) in Eq.(\ref{eq2}) is simply the
Doppler
coupling\cite{D.H.Lee_CSB,Herbut_CSB,Tesanovic_CSB,Ye_CSB}.
$\lambda_{\mu}$ is a Lagrange multiplier to impose the constraint
for the Berry gauge field in the Anderson gauge\cite{Ye_CSB}. The
constraint is that the flux of the Berry gauge field (i.e.,
$\partial \times a^\psi$) is one half of the spinon pair vortex
current (i.e., $\frac{1}{2} j_s^V$) with $j_s^V = \partial \times
\partial \phi_{sp}$, the $3$-current of the spinon pair
vortex\cite{GAUGE_FIELD2}. The factor of $\frac{1}{2}$ represents
the fact that the Dirac spinon acquires the Aharonov-Bohm phase
$\pi$ when it moves around a spinon pair vortex while the phase of
spinon pair field changes by $2\pi$ around the vortex. $A_{\mu}$
is the external electromagnetic field\cite{typeII_CSB}.
$\bar{\rho}_{sp}$ and $\bar{\rho}_{bp}$ are the average density of
the spinon pair and the holon pair respectively. The above low
energy Lagrangian has a local
$U_{a}(1){\times}U_{a^\psi}(1){\times}U_A(1)$ symmetry, which
recovers the local $U_{a^\psi}(1){\times}U_A(1)$ symmetry of the
spinon sector leading to the low energy $d-wave$ Lagrangian if we
replace $a_{\mu}$ with $A_{\mu}$\cite{Ye_CSB} in the spinon
Lagrangian ${\cal L}_{s}$ of Eq.(\ref{eq2}).

Performing the usual duality
transformation\cite{NaLee_CSB,Nayak_V_CSB,Nayak_CSB,Ye_CSB,Savit_CSB,Na1_CSB}
on the first two terms in both ${\cal L}_{s}$ and ${\cal
L}_{b}$\cite{DERIV0} and defining $a_{\mu}^{\Phi_{sp}} =
\lambda_{\mu}/2$, we obtain a dual (vortex) Lagrangian describing
the spinon pair vortex field, $\Phi_{sp}$ and the holon pair
vortex field, $\Phi_{bp}$, \bqa &&Z = \int
{D\psi_l}{D\psi_b}{D\Phi_{sp}}{D\Phi_{bp}}{Dc_{sp\mu}}{Dc_{bp\mu}}{Da_{\mu}^\psi}
{Da_{\mu}^{\Phi_{sp}}}{Da_{\mu}}\nn&& exp\Bigl({-\int{d^3x}{\cal
L} } \Bigr), \nn &&{\cal L} = {\cal L}_{s} + {\cal L}_{b} , \nn
&&{\cal L}_{s} = \bar\psi_l\gamma_\mu(\partial_\mu -
ia_{\mu}^\psi)\psi_l \nn && + |(\partial_\mu -
ia_{\mu}^{\Phi_{sp}} - ic_{sp\mu})\Phi_{sp}|^2 + V(|\Phi_{sp}|) +
\frac{1}{2K_{sp}}|\partial\times{c}_{sp}|^2 \nn && +
i2a_{\mu}^\psi(\partial\times{a}^{\Phi_{sp}})_\mu - i(2a +
K_{sp}^{-1}J_{c})_{\mu}(\partial\times{c}_{sp})_{\mu}\nn &&-
\mu_{sp}(\partial\times{c}_{sp})_{\tau} , \nn &&{\cal L}_{b} =
|\partial_{\mu}\psi_b|^2 + |(\partial_\mu -
ic_{bp\mu})\Phi_{bp}|^2 + V(|\Phi_{bp}|) \nn && +
\frac{1}{2K_{bp}}|\partial\times{c}_{bp}|^2 - i(2a + 2A +
K^{-1}_{bp}J_{b})_{\mu}(\partial\times{c}_{bp})_{\mu} \nn && -
\mu_{bp}(\partial\times{c}_{bp})_{\tau} \label{eq3} \eqa with the
effective Ginzburg-Landau potentials of the spinon (holon) pair
vortex fields $V(|\Phi_{sp(bp)}|) =
m_{\Phi_{sp(bp)}}^2|\Phi_{sp(bp)}|^2 +
\frac{u_{\Phi_{sp(bp)}}}{2}|\Phi_{sp(bp)}|^4$ and
$\mu_{sp(bp)}=\frac{\bar \rho_{sp(bp)}}{K_{sp(bp)}}$\cite{DERIV0}.
Here $\Phi_{sp(bp)}$ is the vortex field of the spinon (holon)
pair and $c_{sp(bp)\mu}$, the vortex (dual) gauge field (or the
spin wave) which mediates interactions between the spinon (holon)
pair vortex fields. It is noted that $a_{\mu}^\psi$ and
$a_{\mu}^{\Phi_{sp}}$ represent the Berry gauge field and the
Lagrangian multiplier field respectively. They are called the
mutual Chern-Simons gauge fields because they guarantee the
mutuality (in Aharonov-Bohm phase contributions) that occurs when
a spinon $\psi_{l}$ moves around a spinon pair vortex $\Phi_{sp}$
or vice versa\cite{Balent99_CSB,Ye_CSB}. It is noted that in our
theory the effect of hole doping into the Mott insulator is
realized in the holon pair vortices which is, in turn, originated
from the holon-pair boson field. The  mathematical form of
Lagrangian ${\cal L}_{s}$ in Eq.(\ref{eq3}) is equivalent to the
$d-wave$ field theory of other
researchers\cite{Tesanovic_CSB,Ye_CSB}. In our theory an
additional part ${\cal L}_{b}$ appears to allow for studies of
doping dependent superconductivity involved with the doped Mott
insulator.

Based on the dual Lagrangian we obtain two different phases in the
underdoped region : (1) the superconducting (SC) phase for $\delta
> \delta_c$ with $\delta_c$, the critical hole concentration for
superconductivity at $0 K$ in the absence of both the spinon pair
vortex condensation and the holon pair vortex condensation
$<\Phi_{sp}> = 0$ and $<\Phi_{bp}> = 0$ and (2) the spin gap (SG)
phase for $\delta < \delta_c$ in the absence of the spinon pair
vortex condensation $<\Phi_{sp}> = 0$ but with the presence of the
holon pair vortex condensation $<\Phi_{bp}> \not= 0$. These two
phases in the vortex field representation are essentially
equivalent to the U(1) holon-pair boson theory of Lee and
Salk\cite{LeeSalk_CSB} in the Higgs field representation.

Integrating over the internal U(1) gauge field $a_\mu$, we obtain
the constraint \bqa
\partial\times{c}_{sp} + \partial\times{c}_{bp} = 0.
\label{eq4} \eqa We insert $c_{bp} = \partial\varphi - c_{sp}$
with $\varphi$, an arbitrary function into Eq.(\ref{eq3}) and
perform a gauge transformation of $\Phi_{bp}\rightarrow
e^{i\varphi} \Phi_{bp}$ to obtain, \bqa &&Z =  \int
{D\psi_l}{D\psi_b}{D\Phi_{sp}}{D\Phi_{bp}}{Dc_{sp\mu}}{Da_{\mu}^\psi}{Da_{\mu}^{\Phi_{sp}}}
e^{-\int{d^3x}{\cal L} } , \nn &&{\cal L} =
\bar\psi_l\gamma_\mu(\partial_\mu - ia_{\mu}^\psi)\psi_l +
|\partial_{\mu}\psi_b|^2 \nn && + |(\partial_\mu -
ia_{\mu}^{\Phi_{sp}} - ic_{sp\mu})\Phi_{sp}|^2 + V(|\Phi_{sp}|) +
\frac{1}{2K_p}|\partial\times{c}_{sp}|^2 \nn && +
i2a_{\mu}^\psi(\partial\times{a}^{\Phi_{sp}})_\mu -
iK_{sp}^{-1}J_{c\mu}(\partial\times{c}_{sp})_{\mu} \nn && +
|(\partial_\mu + ic_{sp\mu})\Phi_{bp}|^2 + V(|\Phi_{bp}|) \nn && +
i(2A + K_{bp}^{-1}J_{b})_{\mu}(\partial\times{c}_{sp})_{\mu} -
\tilde{\mu}(\partial\times{c}_{sp})_{\tau}, \label{eq5} \eqa where
$K_{p}^{-1} = K_{sp}^{-1} + K_{bp}^{-1}$ and $\tilde{\mu} =
\mu_{sp} - \mu_{bp}$. The spinon pair vortex is coupled with the
holon pair vortex via the dual gauge bosons $c_{sp\mu}$.

Redefining $a_{\mu}^{\Phi_{sp}} + c_{sp\mu}$ as
$a_{\mu}^{\Phi_{sp}}$\cite{Ye_CSB}, we obtain the effective dual
Lagrangian \bqa &&Z = \int
{D\psi_l}{D\psi_b}{D\Phi_{sp}}{D\Phi_{bp}}{Dc_{sp\mu}}{Da_{\mu}^\psi}{Da_{\mu}^{\Phi_{sp}}}
e^{-\int{d^3x}{\cal L} } , \nn &&{\cal L} =
\bar\psi_l\gamma_\mu(\partial_\mu - ia_{\mu}^\psi)\psi_l +
i2a_{\mu}^\psi(\partial\times{a}^{\Phi_{sp}})_\mu +
|\partial_{\mu}\psi_b|^2 \nn && + |(\partial_\mu -
ia_{\mu}^{\Phi_{sp}})\Phi_{sp}|^2 + V(|\Phi_{sp}|) \nn && +
|(\partial_\mu + ic_{sp\mu})\Phi_{bp}|^2 + V(|\Phi_{bp}|) +
\frac{1}{2K_p}|\partial\times{c}_{sp}|^2 \nn && - i(2a^{\psi} - 2A
+ K_{sp}^{-1}J_{c} -
K_{bp}^{-1}J_{b})_{\mu}(\partial\times{c}_{sp})_{\mu} \nn && -
\tilde{\mu}(\partial\times{c}_{sp})_{\tau}. \label{eq6} \eqa
Neglecting the contribution of the holon sector of vortex field
($|(\partial_{\mu} + ic_{sp\mu})\Phi_{bp}|^2 + V(|\Phi_{bp}|)$) in
this dual Lagrangian, we recover a mathematically identical form
(but not physics) to the Lagrangian of Ye\cite{Ye_CSB}. Here we
analyze the above Lagrangian in terms of the holon and spinon
vortex fields. In the underdoped region it is expected that phase
coherence of the spinon pairing field appears owing to the large
phase stiffness of spinon pairing field ($K_{sp}$). In the
coherent phase of spinon pairing, the spinon pair vortices form a
neutral dipole pair with its antivortex and thus they do not
contribute vortex condensation. In this case the Berry gauge field
$a_{\mu}^{\psi}$ becomes massive owing to the Anderson-Higgs
mechanism\cite{GAUGE_MASS}. The mass of the Berry gauge field
$a_{\mu}^{\psi}$ is obtained to be proportional to the phase
stiffness of the spinon pairing (i.e., $K_{sp} \sim J
|\Delta_{sp}|^2$)\cite{GAUGE_MASS}. Thus the mass of the Berry
gauge field becomes large in the underdoped region owing to the
large spin gap. With the large mass of the Berry gauge field
$a_{\mu}^{\psi}$, the chiral symmetry breaking is not expected to
occur in the underdoped region\cite{Franz_CSB}. This is because
the fluctuations of the Berry gauge field are suppressed and the
spinons become free in the low energy limit. In such case the
spinon-antispinon pair condensation is not allowed. Thus there can
be no antiferromagnetic order. This is inconsistent with
experiments\cite{Sonier_CSB,Keimer_CSB,Mook_CSB,Niedermayer_CSB,Julien_CSB}.
To properly describe the presence of the antiferromagnetic order
in the underdoped region, we will introduce a dual gauge
transformation involved with the Cooper pair and chargeon pair
fields, which naturally arise as a result of the internal gauge
field $a_\mu$ strongly coupled to the holon pair and spinon pair
fields.

In order to manifestly describe the coexistence of AF and SC
observed in
experiments\cite{Sonier_CSB,Keimer_CSB,Mook_CSB,Niedermayer_CSB,Julien_CSB}
we now introduce two composite vortex pair
fields\cite{Nayak_V_CSB,Nayak_CSB}; the chargeon pair vortex field
$\Psi_{c} =
\Phi_{bp}^{\dagger\frac{1}{2}}\Phi_{sp}^{\dagger\frac{1}{2}}$ and
the Cooper pair vortex field $\Psi_{n} =
\Phi_{bp}^{\frac{1}{2}}\Phi^{\dagger\frac{1}{2}}_{sp}$. $\Psi_{c}$
carries the $-hc/4\tilde{e}$ internal U(1) flux corresponding to
$-4\tilde{e}$ internal U(1) charge while $\Psi_{n}$ carries no
internal U(1) flux. Here $\tilde e$ denotes the unit internal U(1)
gauge charge and the subscript $n$ stands for no internal flux. In
this respect we can also call $\Psi_{c}$ as the "charged" (with
internal U(1) flux of $-\frac{hc}{4 \tilde e}$) vortex field and
$\Psi_{n}$ as the neutral (no internal charge) vortex field. In
the Higgs field language the composite chargeon pair vortex field
$\Psi_{c}$ is originated from the composite nature of the chargeon
pair field defined by $\phi_c = -( \phi_{bp} + \phi_{sp} )$ from
$e^{i\phi_{c}} = e^{-i\phi_{bp}}e^{-i\phi_{sp}}$ and the composite
nature of the Cooper pair vortex field $\Psi_{n}$ comes from the
composite Cooper pair field defined by $\phi_p = \phi_{bp} -
\phi_{sp}$ from $e^{i\phi_{p}} = e^{i\phi_{bp}}e^{-i\phi_{sp}}$.
$\phi_{c}$ carries $-2e$ electromagnetic (EM) charge and
$-4\tilde{e}$ internal gauge charge, while $\phi_{p}$ carries
$+2e$ EM charge and no internal charge. The $+2e$ EM charge of the
Cooper pair as a composite of the holon pair and spinon pair
originates from the $+2e$ EM charge of the holon pair as a
consequence of doped hole charges $+ 2e$. It is of note that for
the hole doped high $T_c$ cuprates the superconducting charge
carriers are known to have electrically positive $+2e$ charges.
For clarity with the terminology here duality between charge and
vortex is summarized in Table [1]. It is of note that our chargeon
pair field is different from the chargon field of the $Z_2$ gauge
theory\cite{SenthilFisher_CSB}; in our slave-boson gauge theory
formulation the chargeon pair field $\phi_c$ carries the U(1)
internal gauge charge $-4\tilde{e}$ and the EM charge $-2e$ as
pointed out above while in the $Z_2$ gauge theory the chargon pair
field carries no internal charge. It carries only the EM charge
$-2e$\cite{SenthilFisher_CSB}. Further our holons are related to
doped holes unlike the case of the $Z_2$ gauge theory.

The spinon pair field and the holon pair field strongly coupled to
the U(1) gauge bosons $a_{\mu}$ lead to the construction of the
composite Cooper pair field, $\phi_{p} = \phi_{bp} - \phi_{sp}$
and the composite chargeon pair field, $\phi_{c} = -(\phi_{bp} +
\phi_{sp})$ respectively. The introduction of these composite
fields are convenient to describe both the superconductivity and
the antiferromagnetism. This is the essence of the present work.
There exists an analogy to quantum chromodynamics (QCD). In QCD
quark-antiquark confinement (i.e., the mesons) exists as low
energy elementary excitations. In a loose sense, the meson
corresponds to the Cooper pair, that is, the confinement of the
spinon pair and the holon pair. Introducing the phase of the
Cooper pair, $\phi_{p} = \phi_{bp} - \phi_{sp}$ and the phase of
the chargeon pair, $\phi_{c} = -(\phi_{bp} + \phi_{sp})$
respectively, we can rewrite Eq.(\ref{eq2}) in terms of these
composite fields and obtain \bqa &&Z = \int
{D\psi_l}{D\psi_b}{D\phi_{p}}{D\phi_{c}}{Da_{\mu}^\psi}{Da_{\mu}^{\Phi_{sp}}}
{Da_{\mu}} e^{-\int{d^3x} {\cal L} } , \nn &&{\cal L} =
\frac{\kappa_{p}}{2}|\partial_\mu\phi_p - 2A_\mu|^2 - i
\bar{\rho}_{sp}( \partial_{\tau}\phi_{p} - A_\tau)  \nn && +
\frac{\kappa_{c}}{2}|\partial_\mu\phi_c + 4a_\mu + 2A_\mu|^2  \nn
&& + \bar{\psi}_l\gamma_\mu (\partial_\mu - ia^{\psi}_\mu)\psi_l
\nn && + \kappa_{cp}(\partial_\mu\phi_c + 4a_\mu +
2A_\mu)(\partial_\mu\phi_p - 2A_\mu) \nn && + i 2
a_{\mu}^{\Phi_{sp}} \left( \partial \times ( a^{\psi} +
\frac{1}{4} \partial (\phi_{p} + \phi_{c}) )\right)_{\mu}  \nn &&
+\frac{1}{2}( \partial_\mu \phi_c + 4a_\mu + 2A_\mu) (J_{c\mu} +
J_{b\mu}) \nn && +\frac{1}{2}( \partial_\mu \phi_p - 2A_\mu)
(J_{c\mu} - J_{b\mu}) \nn && + \frac{J_{c}^2}{2K_{sp}} +
\frac{J_{b}^2}{2K_{bp}} + |\partial_{\mu}\psi_b|^2, \label{eq7}
\eqa where $\kappa_c = \kappa_p = \frac{ {K}_{sp} + {K}_{bp}}{4}$
and $\kappa_{cp} = \frac{ {K}_{sp} - {K}_{bp}}{4}$. As will be
discussed later, in this representation the Cooper pair field
sector in the derived effective Lagrangian describes the
superconductivity while the chargeon pair field sector describes
the antiferromagnetism in the presence of couplings with the Dirac
fermions via the Berry gauge field. From the duality
transformation we obtain an effective field theory to describe the
Cooper pair vortex field and the chargeon pair vortex field.

We write down the effective dual Lagrangian, Eq.(\ref{eq7}) (or
Eq.(\ref{eq6})) in terms of the new composite vortex fields by
redefining $c_{c\mu} \equiv \frac{1}{2}(a_{\mu}^{\Phi_{sp}} -
c_{sp\mu})$ for the chargeon pair dual (vortex gauge) field
associated with the internally `charged' chargeon pair $\phi_c$
with the internal charge $-4\tilde{e}$ and $c_{n\mu} \equiv
\frac{1}{2}(a_{\mu}^{\Phi_{sp}} + c_{sp\mu})$ for the Cooper pair
dual (vortex gauge) field associated with the `neutral' (internal
U(1) charge $0$) Cooper pair $\phi_p$, \bqa &&Z = \int
{D\psi_l}{D\psi_b}{D\Psi_n}{D\Psi_c}{Da_{\mu}^{\psi}}{Dc_{n\mu}}{Dc_{c\mu}}
e^{-\int{d^3x}{\cal L} } , \nn &&{\cal L}  = {\cal L_D} + {\cal
L_O} + {\cal L_{C}} , \nn &&{\cal L_D} =
\bar\psi_l\gamma_\mu(\partial_\mu - ia_{\mu}^\psi)\psi_l +
i4a_{\mu}^\psi(\partial\times{c_c})_\mu + |\partial_{\mu}\psi_b|^2
\nn && + |(\partial_\mu + ic_{c\mu})\Psi_c|^2 + V(|\Psi_c|) +
\frac{1}{2{K}_c}|\partial\times{c}_c|^2 \nn && - {i}(2A
-K_{sp}^{-1}J_{c} + K_{bp}^{-1}J_b)\cdot\partial\times{c_c} +
\tilde{\mu}(\partial\times{c}_c)_{\tau} , \nn &&{\cal L_O} =
|(\partial_\mu + ic_{n\mu})\Psi_n|^2 + V(|\Psi_n|) +
\frac{1}{2{K}_p}|\partial\times{c}_n|^2 \nn && + {i}(2A -
K_{sp}^{-1}J_{c} + K_{bp}^{-1}J_b)\cdot\partial\times{c_n} -
{\tilde{\mu}}(\partial\times{c}_n)_{\tau}, \nn &&{\cal L_{C}} = -
\frac{1}{{K}_{cn}}(\partial\times{c}_c)\cdot(\partial\times{c}_n)
+ {\cal L}_{int}[\Psi_c,\Psi_n], \nn &&{\cal L}_{int} =
\kappa\Bigl(\Psi_{c}^{\dagger}(\partial_\mu\ + ic_{c\mu})\Psi_{c}
- \Psi_{c}(\partial_\mu\ -
ic_{c\mu})\Psi_{c}^{\dagger}\Bigr)\nn&&\times\Bigl(\Psi_{n}^{\dagger}
(\partial_\mu\ + ic_{n\mu})\Psi_{n} - \Psi_{n}(\partial_\mu\ -
ic_{n\mu})\Psi_{n}^{\dagger}\Bigr), \label{eq8} \eqa with ${K}_c =
K_p = \frac{K_{bp}K_{sp}}{K_{bp} + K_{sp}}$, the effective phase
stiffness of each composite Higgs field, $K_{cn} = K_{p}$, the
coupling strength between the supercurrents and $\kappa$, that
between the vortex currents. $V(|\Psi_{c(n)}|) = m_{c(n)}^2
|\Psi_{c(n)}|^2 + \frac{u_{c(n)}}{2}|\Psi_{c(n)}|^2$ is the
Ginzburg-Landau potential of the composite vortex fields
$\Psi_{c}$ and $\Psi_{n}$. The subscript symbols for the
Lagrangians above "D" stand for disordering contribution of the
chargeon pair field; "O", ordering contribution of the Cooper pair
field and "C", coupling between them. It is of note that the
effective phase stiffness is found to be identical for both the
chargeon pair and the Cooper pair, that is, ${K}_c = K_p =
\frac{K_{bp}K_{sp}}{K_{bp} + K_{sp}}$. The phase stiffness
vanishes to zero as hole concentration decreases to
zero\cite{STIFFNESS}. Thus in the low doping region, both the
chargeon pair and the Cooper pair fields are incoherent to
maintain the lowest possible energy state. To put it otherwise,
both the chargeon pair vortex and the Cooper pair vortex are
condensed (i.e., $< \Psi_c > \neq 0$ and $< \Psi_n > \neq 0$
respectively). With increasing hole concentration, the phase
stiffness increases\cite{STIFFNESS}. And thus phase transitions
can occur at a critical hole doping from disordered phases in the
lower doping region to coherent phases in the higher doping region
for both the chargeon pair and Cooper pair fields. Although the
phase stiffness of the chargeon pair is identical to that of the
Cooper pair field, $K_p = K_c$ which varies with hole
concentration, the phase transitions concerned with the chargeon
pair and the Cooper pair do not occur at the same hole
concentration. This is because only the chargeon pair is coupled
to the massless internal U(1) gauge field $a_\mu^\psi$ (the second
term of ${\cal L}_D$ in Eq.(\ref{eq8})) via the chargeon vortex
gauge field $c_{c\mu}$ while the Cooper pair is not, since it is
an internal-charge neutral object\cite{NEUTRALITY_COOPER}. Thus,
despite the identical phase stiffness, the fluctuations of the
gauge field $a^\psi_\mu$ tend to destroy the phase coherence of
the chargeon pair but not that of the Cooper pair since the latter
is not coupled to the gauge field\cite{Fradkin_CSB,Senthil_CSB}.
This tendency is well manifested in the phase diagram of the
Abelian-Higgs model\cite{Fradkin_CSB,Senthil_CSB,Sudbo_CSB} : the
critical value of phase stiffness for the phase transition from an
incoherent phase to a coherent phase increases with increasing
coupling to the gauge field. Applying this result to the chargeon
pair and the Cooper pair, we obtain the following result. The
coherent phase occurs at a larger value of the phase stiffness for
the chargeon pair (which is coupled to the gauge field
$a_\mu^\psi$) than for the Cooper pair (which is not coupled to
the gauge field). Thus for some values of phase stiffness, only
the Cooper pair becomes coherent while the chargeon pair remains
incoherent. This implies that there exists a range of hole
concentration where only the chargeon pair vortices are condensed
without Cooper pair vortex condensation. This corresponds to the
SC phase accompanying the chargeon pair vortex condensation. As
will be shown below, the chargeon pair vortex condensation causes
the antiferromagnetic order which coexists with the
superconductivity.

The dual vortex Lagrangian Eq.(\ref{eq8}) is a different
representation of the Lagrangian Eq.(\ref{eq6}), the former of
which is described by the composite Cooper pair and chargeon pair
vortex fields, $\Psi_{c} = \Phi_{bp}^{\dagger
\frac{1}{2}}\Phi_{sp}^{\dagger \frac{1}{2}}$ and $\Psi_{n} =
\Phi_{bp}^{\frac{1}{2}}\Phi_{sp}^{\dagger\frac{1}{2}}$
respectively. In a different physical context a mathematically
similar procedure is made in the dual vortex theory of Balents et.
al\cite{Nayak_V_CSB,Nayak_CSB} and the SU(2) slave boson theory of
Nagaosa and Lee dealing with instantons\cite{NaLee_CSB}. The Dirac
fermion couples only to the chargeon pair with internally
`charged' vortex $\Psi_{c}$ via the U(1) Berry gauge field
$a_{\mu}^{\psi}$. The massless Dirac fermion acquires the
Aharonov-Bohm phase when it moves around the `charged' (internal
flux $-\frac{hc}{4 \tilde e}$) vortex $\Psi_{c}$, but not so
around the Cooper pair (internal flux $0$) vortex field. This is
because the Cooper pair vortex field cannot carry the internal
U(1) flux\cite{NEUTRALITY_COOPER}. Thus the spinon field is
coupled only to the chargeon pair vortex field via the Berry gauge
field $a_{\mu}^{\psi}$ and the chargeon pair vortex gauge field
$c_{c\mu}$. We note that owing to the $-hc/4\tilde{e}$ internal
U(1) flux the coupling constant in the second term of ${\cal L_D}$
is $-4\tilde{e}$ (with $\tilde{e}$, one unit). As we shall see
below, the Lagrangian ${\cal L_O}$ for the Cooper pair vortex
field is decoupled with ${\cal L_D}$ for the charged vortex sector
in the low energy limit. This is because the Cooper pair field
does not carry the U(1) internal charge and the rest of the
coupling terms vanish in the low energy limit.

Now we examine a possible ground state of chargeon pair from
${\cal L}_D$ of Eq.(\ref{eq8}). Integrating out the Dirac fermion
in the chiral symmetry broken phase, ${\cal L}_D$ leads to the
dual Lagrangian of the Abelian Higgs model for the chargeon pair
field $\phi_c$. It is recalled that the chargeon pair has internal
charge $-4\tilde{e}$ and electromagnetic charge $-2e$. It is known
that such Lagrangian shows a confinement-deconfinement phase
transition of the internal charge $\tilde{e}$ depending on the
phase stiffness $K_c$ of the chargeon pair
field\cite{NaLee_CSB,Fradkin_CSB,Senthil_CSB,Sudbo_CSB}; (1) the
deconfinement phase of internal charge for large phase stiffness
and (2) the confinement phase of internal charge for small phase
stiffness. In the deconfinement phase of the internal charge
$\tilde e$, one can expect an elementary excitation carrying an
internal charge $\tilde{e}$ and fractional electric charge
$\frac{e}{2}$\cite{Sudbo_CSB}. Here we considered only the
confinement phase of the internal charge. This is justifiable in
the low doping region of present interest because the phase
stiffness of chargeon pair ${K}_c = \frac{K_{bp}K_{sp}}{K_{bp} +
K_{sp}}$ is small for low hole doping. As is shown earlier, the
phase stiffness ${K}_c = \frac{K_{bp}K_{sp}}{K_{bp} + K_{sp}}$
diminishes as hole concentration decreases (to zero) at
half-filling owing to the decreasing trend of phase stiffness of
holon pair $K_{bp}$. Thus the chargeon pair vortices are condensed
owing to the small phase stiffness in the low doping region. In
this confinement phase no elementary excitations with the internal
charge $\tilde{e}$ and the fractional electromagnetic charge of
$e$ appear. In the following, we consider the dynamics of Dirac
fermions in the confinement phase of internal charge $\tilde{e}$.

Integrating over the dual gauge field $c_{c\mu}$ in the presence
of the chargeon pair vortex condensation, i.e., $<\Psi_{c}> =
\bar{\Psi}_{c} \not= 0$, we obtain the low energy effective
Lagrangian \bqa &&Z = \int
{D\psi_l}{D\psi_b}{D\Psi_n}{Da_{\mu}^{\psi}}{Dc_{n\mu}}
e^{-\int{d^3x}{\cal L} } , \nn &&{\cal L}  = {\cal L_D} + {\cal
L_O} , \nn &&{\cal L_D} = \bar\psi_l\gamma_\mu(\partial_\mu -
ia_{\mu}^\psi)\psi_l + |\partial_{\mu}\psi_b|^2 +
\frac{1}{4\bar{\Psi}_c^2}\Bigl(4\partial\times{a}^\psi \nn && -
2\partial\times{A} + K_{sp}^{-1}\partial\times{J_{c}} -
K_{bp}^{-1}\partial\times{J_b}
 + i{K}_p^{-1}\partial\times\partial\times{c}_n \nn &&
-2\kappa\bar{\Psi}_{c}^2\Bigl(\Psi_{n}^*(\partial + ic_n)\Psi_{n}
- \Psi_{n}(\partial - ic_n)\Psi_{n}^*\Bigr)\Bigr)^2 , \nn &&{\cal
L_O} = |(\partial_\mu + ic_{n\mu})\Psi_n|^2 + V(|\Psi_n|) +
\frac{1}{2{K}_p}|\partial\times{c}_n|^2 \nn && + i(2A -
K_{sp}^{-1}J_{c} + K_{bp}^{-1}{J_b})\cdot\partial\times{c_n} -
\tilde{\mu}(\partial\times{c}_n)_{\tau}. \label{eq9} \eqa We
stress that the two effective Lagrangians above contain essential
physics. The total effective Lagrangian ${\cal L}$ is separated
into two parts; (1) ${\cal L_D}$ for the chargeon pair vortex
field and (2) ${\cal L_O}$ for the Cooper pair vortex field.
${\cal L_D}$ also describes the dynamics of the holon
quasiparticles effectively coupled to the Dirac spinons in the
ground state of randomly disordered chargeon pair including the
self-interacting Dirac fermions coupled to the massless gauge
bosons $a_\mu^\psi$. The condensation of the chargeon pair vortex
field makes the Berry gauge field ${a}^{\psi}$ remain massless
even in the superconducting state as discussed
earlier\cite{MASS2}. This is different from other
theories\cite{Herbut_CSB,Tesanovic_CSB,Ye_CSB,SU2_CSB} in which
the Berry gauge field becomes massive in the superconducting
state. In these theories\cite{Herbut_CSB,Tesanovic_CSB} the Berry
gauge field is coupled to the Cooper pair field. The Berry gauge
field becomes massive in the superconducting phase owing to the
Anderson-Higgs mechanism. Accordingly the Dirac fermion ends up
being coupled to the massive Berry gauge field in the
superconducting state. On the other hand, in our theory the Berry
gauge field is coupled to the chargeon pair field, but not to the
Cooper pair field. Thus the Berry gauge field remains massless
even in the superconducting state accompanying the condensation of
the chargeon pair vortex. As a result the Dirac fermion and the
massless Berry gauge field is described by the QED Lagrangian. The
chiral symmetry breaking occurs since the flavor number of
fermions of our interest is $2$ which is less than the critical
flavor number $N_c = 3.24$\cite{CSB_DETAIL}. The chiral symmetry
breaking results in the antiferromagnetic order. This, in turn,
implies the coexistence of the AF and the SC because we obtained
the AF in the presence of SC accompanying the condensation of the
chargeon pair vortex. If we do not introduce the vortex pair field
variables, $\Psi_c$ and $\Psi_n$, the SC state is characterized by
the absence of the spinon pair vortex ($<\Phi_{sp}> = 0$) and the
holon pair vortex ($<\Phi_{bp}> = 0$) respectively in the original
representation. With this original representation, we can not
address the antiferromagnetism. The reason is as follows. The
Berry gauge field becomes massive in the presence of coherent
spinon pairing owing to the Anderson-Higgs
mechanism\cite{GAUGE_MASS}. Thus the chiral symmetry breaking does
not occur owing to the large mass of the Berry gauge field in the
low doping region. Accordingly the antiferromagnetic order does
not occur. To properly describe the antiferromagnetic order in the
underdoped region, it is, thus, shown to be necessary to introduce
the composite pair fields, namely the Cooper pair field and the
chargeon pair field respectively. This is because the strong gauge
fluctuations allow the spinon pair field and the holon pair field
to form the composite pair fields. As will be soon be seen below,
the effective dual Lagrangian for the Cooper pair vortex and
chargeon pair vortex fields is a convenient `tool' to directly see
whether the coexistence of the AF and the SC affects the
superconductivity. In this pair vortex field representation it is
natural to see the chargeon pair vortex condensation because of
the small phase stiffness
$K_{c}$\cite{NaLee_CSB,Fradkin_CSB,Senthil_CSB,Sudbo_CSB} for the
case of low hole doping in the underdoped region, thus causing the
massless Berry gauge field in the superconducting
state\cite{MASS2}. The Berry gauge field $a_{\mu}^{\psi}$ has the
kinetic energy which varies with the coupling strength
proportional to the square of the charged vortex field
$\bar{\Psi}_{c}^2$.

In order to examine whether the massless nodal fermion excitations
affect the condensation of the Cooper pair vortex field in the
presence of chargeon pair vortex condensation, we rearrange the
above effective Lagrangian as \bqa &&{\cal L}_{AF} =
\bar\psi_l\gamma_\mu(\partial_\mu - ia_{\mu}^\psi)\psi_l  , \nn
&&{\cal L}_{SC} = |(\partial_\mu + ic_{n\mu})\Psi_n|^2 +
V(|\Psi_n|) + \frac{1}{2{K}_p}|\partial\times{c}_n|^2 \nn && + i
2A \cdot\partial\times{c_n} -
\tilde{\mu}(\partial\times{c}_n)_{\tau}, \nn &&{\cal L}_{C} =
\frac{1}{4\bar{\Psi}_c^2}\Bigl(4\partial\times{a}^\psi -
2\partial\times{A} \nn && + K_{sp}^{-1}\partial\times{J_{c}} -
K_{bp}^{-1}\partial\times{J_b} +
i{K}_p^{-1}\partial\times\partial\times{c}_n \nn &&
-2\kappa\bar{\Psi}_{c}^2\Bigl(\Psi_{n}^*(\partial + ic_n)\Psi_{n}
- \Psi_{n}(\partial - ic_n)\Psi_{n}^*\Bigr)\Bigr)^2  \nn && - i(
K_{sp}^{-1}J_{c} - K_{bp}^{-1}{J_b})\cdot\partial\times{c_n}.
\label{eq10} \eqa Here coupling between the SC and the AF are in
${\cal L}_{C}$. We can judge whether the coupling terms are
relevant for SC phase transition, by determining the scaling
dimensions of coupling constants in the renormalization group
procedure. From a dimensional analysis\cite{SCALING_DIMENSION},
the scaling dimensions for all of the coupling constants in ${\cal
L}_{C}$ are obtained to be negative, that is, $\left[
\frac{1}{\Psi_c^2} \right] = -1$, $\left[ K_{sp}^{-1} \right] =
-1$, $\left[ K_{bp}^{-1} \right] = -1$, $\left[ K_{p}^{-1} \right]
= -1$ and $\left[ \kappa  \right] = -1$. Since the scaling
dimension of coupling constants are negative, the strength of
coupling diminishes as the cut-off energy gets smaller by
integrating out high momentum modes in the renormalization group
procedure\cite{CHAIKIN}. Thus all of the coupling terms in ${\cal
L}_{C}$ vanish in the low energy limit. Physically speaking,
antiferromagnetism (${\cal L}_{AF}$) and superconductivity (${\cal
L}_{SC}$) are decoupled in the low energy limit. As a consequence
the antiferromagnetism can not affect the superconducting phase
transition at $T = 0$, thus allowing the XY universality class in
the extreme type II limit\cite{Kim_CSB}. We note that our
superconducting phase preserves the local
$U_{a}(1)\times{U}_{a^\psi}(1)$ gauge symmetry. Only $U_A(1)$
symmetry is broken. In this sense our $d-wave$ superconducting
state is conventional.

It is now obvious from the examination of ${\cal L}_{AF}$ in
Eq.(\ref{eq10}) that the dynamical mass generation for the Dirac
fermion leads to an antiferromagnetic
ordering\cite{DonKim_CSB,Herbut_CSB,Marston_CSB} since the chiral
symmetry associated with the translational symmetry in this spinor
representation\cite{Herbut_CSB} is broken\cite{CSB_DETAIL}. The
$d-wave$ superconductivity can coexist with the spin density wave
(SDW) order. Magnetization is shown to be proportional to the
amplitude of the chargeon pair vortex condensation (coupling
strength between the nodal spinon and the Berry gauge field),
i.e., $\bar{\Psi}_{c}^{2}$. The vortex condensation amplitude is
given by the stiffness parameter, $\bar{\Psi}_{c}^{2} \sim
(\bar{K}_{c} - K_{c})$. Here $\bar{K}_{c}$ is the critical
stiffness parameter where the chargeon pair vortices begin to
condense. As discussed earlier, for the case of $K_c >
\bar{K}_{c}$ where the chargeon pair vortices are not condensed,
deconfinement of the internal U(1) gauge charge occurs. We obtain
the magnetization as a function of hole doping, $m \sim
(\bar{K}_{c} - K_{c}) \sim (\delta_{c}^{in2} - \delta^2)$ where
$\delta_{c}^{in}$ is the critical hole concentration corresponding
to the critical phase stiffness $\bar{K}_{c}$. The critical hole
doping $\delta_{c}^{in}$ of the confinement to deconfinement
transition is larger than the critical hole doping $\delta_{c}$ of
the superconducting transition as discussed earlier. To determine
the precise value of $\delta_{c}^{in}$ is beyond the scope of our
paper. As mentioned above the presence of the antiferromagnetic
order does not affect the $d-wave$ superconductivity.

We summarize our results in Table [2]. In the underdoped
superconducting region of $\delta > \delta_c$ with $\delta$, the
hole concentration and $\delta_{c}$, the critical hole
concentration for initiating the SC phase, the chargeon pair
vortex $\Psi_c$ (but not the Cooper pair vortex $\Psi_n$) is
condensed. In the PG phase ($\delta < \delta_c$) the Cooper pair
vortex becomes condensed and, if possible, only a local phase
coherence can survive. This is the pseudogap phase of the
preformed pair scenario\cite{Preformed_CSB}. This vortex
superfluid state corresponds to the Mott insulating phase of the
Cooper pair field\cite{Na1_CSB}. In the chargeon pair vortex field
sector we considered only the chargeon pair vortex condensed phase
$<\Psi_c> \not= 0$ corresponding to the confinement phase of the
internal charge. Owing to the disordered chargeon pairs the Berry
gauge field $a_{\mu}^{\psi}$ remains massless even in the SC
phase\cite{MASS2}. The Berry gauge field $a_{\mu}^{\psi}$
describes the Aharonov-Bohm phase acquired by spinon when it moves
around chargeon vortex. In the SC phase the chiral symmetry
breaking also occurs and thus the Dirac fermions become massive.
The antiferromagnetic order originated from the massive Dirac
fermions near the nodal points emerges in the underdoped $d-wave$
SC state in the low temperature ($T = 0$) limit.

\section{Summary}

In conclusion, we showed that the antiferromagnetic order is
decoupled to the superconducting order in the low energy limit.
Thus the chiral symmetry breaking (the antiferromagnetic order)
does not affect the XY universality class of the extreme type II
superconducting phase transition at $T = 0$. In short the $d-wave$
superconductivity is not affected by the antiferromagnetic order
despite the coexistence between the two. In the present paper, we
concentrated on the dynamics of the Dirac fermion only in the
confinement phase of internal charge $\tilde{e}$. It is of great
interest to study the dynamics of the Dirac fermion in the
deconfinement phase of internal charge $\tilde{e}$ in the future.

\begin{table*}
\caption{\label{Table 1} Duality of charges and vortices}
\begin{tabular}{cccccccc}
\hline
   & Internal charge & Electromagnetic charge \nn
   & or flux quantum  & or flux quantum  \nn
\hline Cooper pair field $e^{i\phi_{p}}$ &  $0$ &  $+2e$    \nn
Cooper pair vortex field $\Psi_{n}$  & $0$ &  $+\frac{hc}{2e}$ \nn
Chargeon pair field $e^{i\phi_{c}}$   &  $-4\tilde{e}$   &  $-2e$
\nn Chargeon pair vortex field  $\Psi_{c}$  &
$-\frac{hc}{4\tilde{e}}$ &  $-\frac{hc}{2e}$\nn \hline
\end{tabular}
\end{table*}

\begin{table*}
\caption{\label{Table 2}Phases of vortex fields for pseudogap
phase and superconducting phase}
\begin{tabular}{cccccccc}
\hline
  & Pseudogap phase &  Superconducting phase  \nn
  & ($\delta<\delta_{c}$)  &  ($\delta_{c}<\delta$ (underdoped))  \nn
\hline
  Phase of vortex field &  $<\Psi_{n}> \not= 0$ and $<\Psi_{c}> \not= 0$   &  $<\Psi_{n}> = 0$ and $<\Psi_{c}> \not= 0$   \nn
  Physical state & MI, AF   &  SC, AF  \nn  \hline
\end{tabular}
\end{table*}


\begin{thebibliography}{9}
\bibitem{CSB_DETAIL}
The ground state of this QED Lagrangian is characterized by the particle-hole pair order parameter, i.e., $<\bar{\psi}_{l}\psi_{l}>$ depending on the flavor number $N$ of the massless Dirac fermions\cite{DonKim_CSB,QED3_CSB,Herbut_CSB,Tesanovic_CSB}.
Beyond the critical flavor number $N_c$, the ground state is simply the particle-hole vacuum (the chiral-symmetry preserved state), i.e., $<\bar{\psi}_{l}\psi_{l}> = 0$.
In this phase the Dirac fermions remain massless.
Below $N_{c}$ the ground state becomes a particle-hole pair condensate (the chiral-symmetry broken state), i.e., $<\bar{\psi}_{l}\psi_{l}> \not= 0$.
In this phase the Dirac fermion becomes massive owing to the binding with its antiparticle.
According to the previous works\cite{QED3_CSB,DonKim_CSB,Herbut_CSB,Tesanovic_CSB}, the critical flavor number is known to be $N_c = 3.24$ in the $1/N$ expansion calculation.
Because of $N = 2 < N_c$ for the nodal Dirac fermions, the dynamical mass generation occurs, i.e., the chiral symmetry is broken to allow the antiferromagnetic order.
\bibitem{GAUGE_FIELD2}
In the Anderson gauge\cite{Ye_CSB}, one chooses
$\phi_{\uparrow} = \phi_{sp}$ and  $\phi_{\downarrow} =0$.
Then the Berry gauge field is obtained to be
$a^{\psi}_{\mu} = \frac{1}{2}\partial_{\mu} \phi_{sp}$ and
the flux of the Berry gauge field is determined by the vortex current of spinon pair
$(\partial \times a^{\psi})_{\mu} = \frac{1}{2} ( \partial \times \partial \phi_{sp} )_{\mu} = \frac{1}{2}j^V_{\mu}$.
This phase is geometric in the sense that the total phase acquired by a spinon is determined only by the number of vortices enclosed by the path of spinon but not by dynamics.
This is why it is called the `Berry' gauge field.
\bibitem{DERIV0}
Here we obtain dual Lagrangian for the spinon pair field and the holon pair field in Eq.(\ref{eq2}),
\bqa
&& {\cal L} = \frac{K}{2} (\partial_{\mu}\phi - A^{eff}_\mu )^2
+ i\bar{\rho} \partial_{\tau}\phi, \nonumber
\eqa
where
$\phi = \phi_{sp}$ ($\phi_{bp}$), the spinon (holon) pair field,
$K = K_{sp}$ ($K_{bp}$) is the phase stiffness of the spinon (holon) pair field,
$\bar {\rho} = \bar{\rho}_{sp}$ ($\bar{\rho}_{bp}$), the density of the spinon (holon) pair,
and $A^{eff}_\mu = 2a_\mu + K_{sp}^{-1}J_{c\mu}$ ($2a_\mu + 2A_\mu + K_{bp}^{-1}J_{b\mu}$), the effective gauge field coupled to the spinon (holon) pair.
After the Hubbard-Stratonovich transformation, one obtains
${\cal L} = \frac{1}{2K} j_{\mu}^2 + i j_\mu ( \partial_{\mu}\phi - A^{eff}_\mu )
+ i\bar{\rho} \partial_{\tau}\phi$.
Rewriting $j_\mu^{'} = ( j_0 + \bar{\rho}, j_1, j_2)$ and integrating over the smooth part of phase fluctuations of the spinon (holon) pair, one obtains
\bqa
&& {\cal L}_{b}  = | (\partial_\mu - ic_{\mu}) \Phi|^2 + V(|\Phi_{bp}|)  \nn
&& + \frac{1}{2K} |\partial\times c|^2 - i A^{eff}_{\mu} (\partial \times c)_{\mu} - \mu(\partial\times{c})_{\tau} \nn
&& +i A^{eff}_\tau \bar{\rho} + \frac{ \bar{\rho}^2}{2K}, \nonumber
\eqa
where $j^{'}_\mu = ( \partial\times{c})_\mu$ and $\mu = \frac{\bar{\rho}}{K}$.
With the neglect of the last two terms, this Lagrangian reduces to the vortex Lagrangian in Eq.(\ref{eq3}).
\bibitem{GAUGE_MASS}
In the coherent phase of spinon pairing, the mass of the Berry gauge field $a_{\mu}^{\psi}$ is generated owing to the Anderson-Higgs mechanism.
To see this we consider the Lagrangian for the spinon pair vortex along with the mutual Chern-Simons term,
\bqa
&&{\cal L} = |(\partial_\mu - ia_{\mu}^{\Phi_{sp}})\Phi_{sp}|^2 + m_{\Phi_{sp}}^2|\Phi_{sp}|^2  \nn
&& + i2a_{\mu}^\psi(\partial\times{a}^{\Phi_{sp}})_\mu
\nonumber
\eqa
in Eq.(\ref{eq6}) with the neglect of the higher order contributions for the potential energy of the spinon pair vortex.
Here the mass of the spinon pair vortex is proportional to the core energy of vortex and thus is proportional to the phase stiffness of the spinon pair, that is, $m_{\Phi_{sp}} \sim K_{sp} \sim J |\Delta_{sp}|^2$.
Integrating over $\Phi_{sp}$ and $a_{\mu}^{\Phi_{sp}}$, one obtains
\bqa
&&{\cal L}^{'} = m_{a^{\psi}}^2 (a^\psi)^2
\nonumber
\eqa
with $m_{a^{\psi}} \sim K_{sp} \sim J |\Delta_{sp}|^2$.
\bibitem{STIFFNESS}
In the holon-pair boson theory, the phase stiffness of the spinon
pair and the holon pair are given by $K_{sp} \sim J_\delta
|\Delta_{sp}|^2$ and $K_{bp} \sim J |\Delta_{bp}|^2
|\Delta_{sp}|^2$ respectively. In the underdoped region, $K_{sp}
>> K_{bp}$ because $|\Delta_{bp}| \approx  \delta$ (at $0K$)
decreases to zero as half-filling is approached (with $\delta$,
the hole concentration). From this we obtain
$\frac{K_{bp}K_{sp}}{K_{bp} + K_{sp}} \approx K_{bp} = J
|\Delta_{sp}|^2 \delta^2$. Thus the phase stiffness of both the
Cooper pair and the chargeon pair increases with increasing hole
concentration in the underdoped region.
\bibitem{NEUTRALITY_COOPER}
In the U(1) holon-pair boson theory\cite{LeeSalk_CSB} the Cooper pair is represented as a composite of the spinon pair and the holon pair, and has no internal gauge charge even though both the spinon pair and the holon pair have internal charge $2\tilde e$.
This is because the Cooper field ($\Delta_{bp} \Delta_{sp}^*$) is invariant under the internal gauge transformation,
$\Delta_{bp}^{'} = e^{i \theta}\Delta_{bp}$ and
$\Delta_{sp}^{'} = e^{i \theta}\Delta_{sp}$.
\bibitem{MASS2}
In the presence of the chargeon pair vortex condensation
$<\Psi_c>=\bar{ \Psi_c}$, there occurs only the Maxwellian action
for the gauge field $a_\mu^\psi$ without the mass term as is shown
in the third term of ${\cal L}_D$ in Eq.(\ref{eq9}). This is
because the Meissner effect does not occur in the presence of the
chargeon pair vortex condensation.
\bibitem{SCALING_DIMENSION}
From a dimensional analysis, the scaling dimensions of field variables are obtained :
$\left[ a_{\mu}^{\psi} \right]   =  1$,
$\left[ A_{\mu} \right]   =  1$,
$\left[ \psi_l \right]    =  1$,
$\left[ c_{n \mu} \right]  =  1$,
$\left[ \Psi_n \right]  =  \frac{1}{2}$ and
$\left[ \psi_b \right]  =  \frac{1}{2}$.
Here we denote the relative scaling dimension of variable $X$ with respect to momentum as $[X]$.
We use the unit system where the Fermi velocity and Planck constant are set to be 1, that is, $v_F = h = 1$.
From the condition that the action should be dimensionless, one can readily calculate the scaling dimension of coupling constants.
The scaling dimensions for all of the coupling constants in ${\cal L}_{ir.}$ are obtained to be negative.
They are
$\left[ \frac{1}{\Psi_c^2} \right] = -1$,
$\left[ K_{sp}^{-1} \right] = -1$,
$\left[ K_{bp}^{-1} \right] = -1$,
$\left[ K_{p}^{-1} \right] = -1$ and
$\left[ \kappa  \right] = -1$.
\end{thebibliography}
\end{document}